\documentclass[aps,prl,twocolumn,showpacs,superscriptaddress]{revtex4}
\usepackage{graphicx}
\newlength{\picwidth}
\begin{document}
\draft

\title{Enhancement of localization in one-dimensional random potentials with
long-range correlations}

\author{U.~Kuhl}
\affiliation{Fachbereich Physik, Philipps-Universit\"{a}t Marburg, Renthof 5,
D-35032 Marburg, Germany}
\author{F.~M.~Izrailev}
\affiliation{Instituto de F\'{\i}sica, Universidad Aut\'{o}noma de
Puebla, Apartado Postal J-48, Puebla, Pue., 72570, M\'{e}xico}

\author{A.~A.~Krokhin}
\affiliation{Instituto de F\'{\i}sica, Universidad Aut\'{o}noma de
Puebla, Apartado Postal J-48, Puebla, Pue., 72570, M\'{e}xico}
\affiliation{Department of Physics, University of North Texas,
P.O. Box 311427, Denton, TX 76203}

\date{\today}

\begin{abstract}
We experimentally study the effect of enhancement of localization in weak
one-dimensional random potentials. Our experimental setup is a single mode
waveguide with 100 tuneable scatterers periodically inserted into the waveguide.
By measuring the amplitudes of transmitted and reflected waves in the
spacing between each pair of scatterers, we observe a strong decrease of the
localization length when white-noise scatterers are replaced by
a correlated arrangement of scatterers.
\end{abstract}


\maketitle

Processes of destructive interference at backscattering do not
vanish after averaging over disorder, unlike interference at
scattering over other directions. In a random potential this
property of backscattering may lead to Anderson
localization \cite{and58}. Experimentally it is difficult to find
Anderson localization for electrons due to Coulomb and electron-phonon interaction,
but it was found for photons in optically diffusive media \cite{gen87,cha00,sch07}
(for a review see Ref. \cite{gen05b}). In case of white-noise one-dimensional
potentials $U(x)$, where backscattering leads to localization at
{\it any} energy $E$ independently of the strength of the potential,
the localization length $l(E)$ is determined by spectral
composition of the correlation function \cite{lif88,mak99b},
\begin{equation} \label{eq:main}
  \lambda(E) \equiv {l^{-1}(E)} = ({\sigma^2}/{8 k^2}) W(2k).
\end{equation}
Here $k^2=E$ is the energy of an eigenstate, $W(2k)$ is the
Fourier harmonic of the correlator $\langle
U(x)U(x^{\prime})\rangle$, and $\sigma^2 = \langle U^2(x) \rangle -
\langle U(x) \rangle^2 $ is the variance of disorder.
Eq.~(\ref{eq:main}) is obtained in the first (Born) approximation
over weak disorder, i.\,e., $\sigma^2 \ll 1$.

A common opinion is that the shortest localization length is reached
for the most disordered (uncorrelated) potentials with white-noise
spectrum $W(k)=1$. This opinion is based on the fact that
correlations, reducing the degree of disorder, typically give
rise to extended states. In the dimer model, short-range
correlations result in two resonant extended states
[\onlinecite{dun90,phi91}], that was observed in a semiconductor
superlattice [\onlinecite{bel99}]. A continuum of extended states
for potentials with long-range correlations was predicted in Ref.
[\onlinecite{mou98,izr99}] and was experimentally verified in
a microwave waveguide with intentionally introduced correlated
disorder [\onlinecite{kuh00a}]. Thus, there is a strong evidence
that correlations may destroy localization \cite{tit05}.

In this Letter we address the opposite situation and show that
correlations in weakly disordered potentials can {\it enhance}
localization for a continuum of states, resulting in the
localization length much shorter than that in a white-noise
potential with the same $\sigma$. The effect of localization
enhancement is important for random lasers \cite{cao99,pat03}, where
extremely localized states provide higher efficiency. Our
results clearly demonstrate that the long-range correlations may
either suppress or enhance localization. This conclusion probably
has to change a common point of view that correlations, being a
manifestation of some kind of order, may only suppress localization.

\begin{figure}[h]
\includegraphics[width=8.5cm]{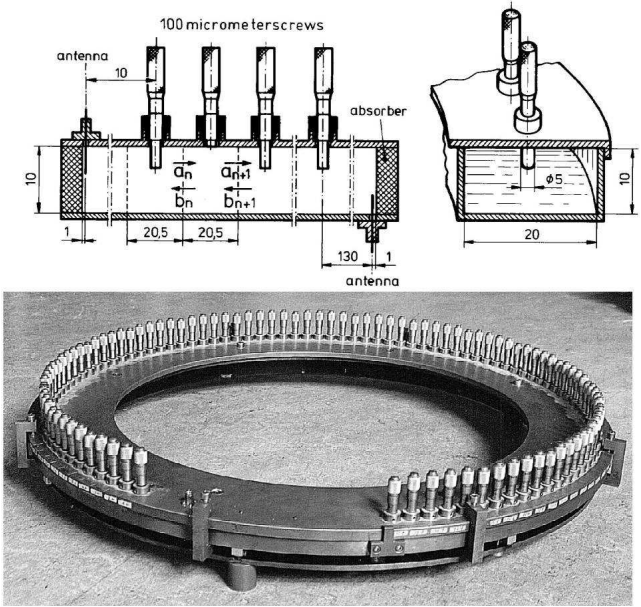}\\
\caption{\label{fig:setup}
Schematic view of the waveguide with 100 scatterers. The waveguide is closed by
microwave absorbers at both ends.
The lower antenna can be placed anywhere within the scattering arrangement.}
\end{figure}

The setup is a single mode waveguide with transverse dimensions
$a=20$\,mm, $b=10$\,mm, and with 100 cylindrical scatterers of
radius $r=2.5$\,mm, periodically inserted with spacing of
$d=20.5$\,mm, see Fig.~\ref{fig:setup}. In the experiment, we have
exploited the transmission and reflection of the lowest waveguide
mode in the frequency range between $\nu_{\rm min}={c}/{2a}
\approx$ 7.5\,GHz and $\nu_{\rm max}={c}/{a}={c}/{2b} \approx$
15\,GHz. The dispersion relation in an empty rectangular microwave
waveguide is given by $k=({2\pi}/{c})\sqrt{\nu^2-\nu^2_{\rm
min}}$, where $c$ is the speed of light. Thus the normalized wave
vector $kd/\pi$ ranges from 0 to 1.8. The setup was already used to show
the ``Hofstadter butterfly'' \cite{kuh98b} and mobility edges
emerging from correlated disorder \cite{kuh00a}. The arrangement
allows to assemble a random potential with prescribed correlation
function.

The strength $U_n$ of a single scatterer is associated with its
length embedded inside the waveguide, and the length is
varied by micrometer screws. When $U_n=const$, the waveguide is
fully transparent in the Bloch band.
For the white-noise potential 100 numbers
$\varepsilon_n =U_n-\langle U_n \rangle$ were drawn as
uncorrelated random numbers with variance $\sigma^2$. The
corresponding localization length was much larger than the length
$L$ of the waveguide, thus the eigenstates are delocalized within
the waveguide. However, if the white-noise potential $\varepsilon_n$
is replaced by a correlated sequence, the localization length inside
a prescribed window of energy is strongly reduced and the effects of
localization in the transmission and reflection become observable.

The experiment can be well-described by a one-dimensional equation
of Kronig-Penney type \cite{kuh00a},
\begin{equation}\label{eq:discrete}
  \psi ^{\prime \prime }(x)+k^{2}\psi (x)=\sum_{n=-\infty }^{\infty }U_{n}\psi (x_{n})\delta (x-x_{n}).
\end{equation}
Here $U_{n}$ is the amplitude of the $n$th delta-scatterer located
at $x=x_{n}=n d$. Therefore Eq.~(\ref{eq:main}), which assumes an
electron with parabolic dispersion, is replaced for an electromagnetic waveguide by \cite{kuh00a}
\begin{equation} \label{eq:locKP}
  l^{-1} (E)=\frac{\sigma^2k^2}{8}\frac{\sin^2(kd)}{\sin ^2\mu }W(2\mu).
\end{equation}
Here $\sigma^2 = \langle U_n^2 \rangle - \langle U_n\rangle^2$. The
phase $\mu $ is given by the Kronig-Penney dispersion
relation,
\begin{equation} \label{eq:disp}
  2\cos \mu=2\cos(kd)+ \frac{\langle U_n \rangle}{k} \sin(kd), \,\,\,\, 0\leq \mu \leq \pi.
\end{equation}
The function $W(2\mu)$ is the Fourier transform of the binary
correlator $\zeta (s)= \langle
\varepsilon_{n}\varepsilon_{n+s}\rangle/ \sigma^{2}$,
\begin{equation} \label{eq:W}
  W(2\mu)=1\;+\;2\sum_{s=1}^{\infty}\zeta (s)\cos(2s\mu).
\end{equation}
It follows from Eq.~(\ref{eq:W}) that $W(2\mu)$ is symmetric with
respect to the band center $\mu = \pi/2$.

\begin{figure}
\includegraphics[width=\picwidth]{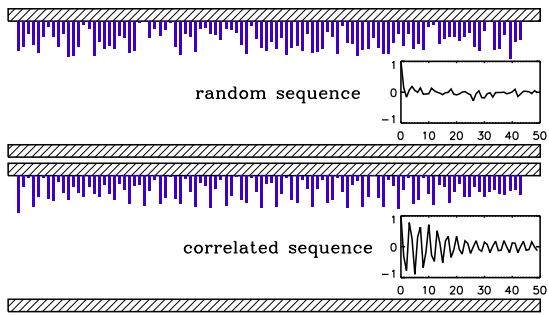}\\
\caption{\label{fig:disorder}
Profile of intrusion of all 100 micrometer screws into
the waveguide for uncorrelated and correlated random sequence $U_n$.
Insets show the corresponding correlation function calculated from the micrometer screw depths
(Eq.~(\ref{eq:corr})).}
\end{figure}

For any white-noise potential the correlator $\zeta(s) \equiv
0$, apart from $\zeta(0) \equiv 1$, leading to $W(2\mu) =1$.
To observe the effect of enhancement
of localization it is necessary to have the function $W(2\mu) =W_0
\gg 1$ within some interval $\Delta \mu = \mu_2 - \mu_1$. Because
of the normalization condition
\begin{equation} \label{eq:norm}
  \zeta(0)=\frac{2}{\pi}\int_{0}^{\pi/2} W(2\mu) d \mu =1,
\end{equation}
the width $\Delta \mu$ and the enhancement factor $W_0$ are related,
$2W_0= \pi\Delta \mu\ $, providing that $W(2\mu)$ vanishes outside
the interval $[\mu_1,\mu_2]$. Thus, the correlation-induced
enhancement of localization within the interval $\Delta \mu$ is
accompanied by full transparency of the waveguide for all other
frequencies \cite{endnote}. The Fourier coefficients of the
step-function, that is $W(2\mu) = W_0$ within the interval
$\Delta \mu$ and $W(2\mu)=0$ otherwise, is given by
\begin{equation}\label{eq:corr}
  \zeta(s)=\frac{1}{2s} \frac{\sin(2s\mu_{2})-\sin(2s\mu_{1})}{\mu_{2}-\mu_{1}}.
\end{equation}
The inverse-power-law decay of $\zeta(s)$ is a signature of the
long-range correlations. A correlated sequence was generated
using the algorithm proposed in Ref.~\cite{kuh00a}. The two random
sets used in the experiments are shown in Fig.~\ref{fig:disorder},
including their correlations.

For both sets the elements $S_{12}=S_{21}^*$ and $S_{22}$ of the
scattering matrix were measured as functions of antenna position.
Here $S_{12}$ ($S_{22}$) is the transmission (reflection) amplitude of the scattering process when
the fixed antenna is in front of the first scatterer ($n=1$) and
the moving antenna is located between the $n$th and $(n+1)$st
scatterer. The moving antenna emits and receives the signal while
measuring $S_{22}$.

\begin{figure}
\includegraphics[width=\picwidth]{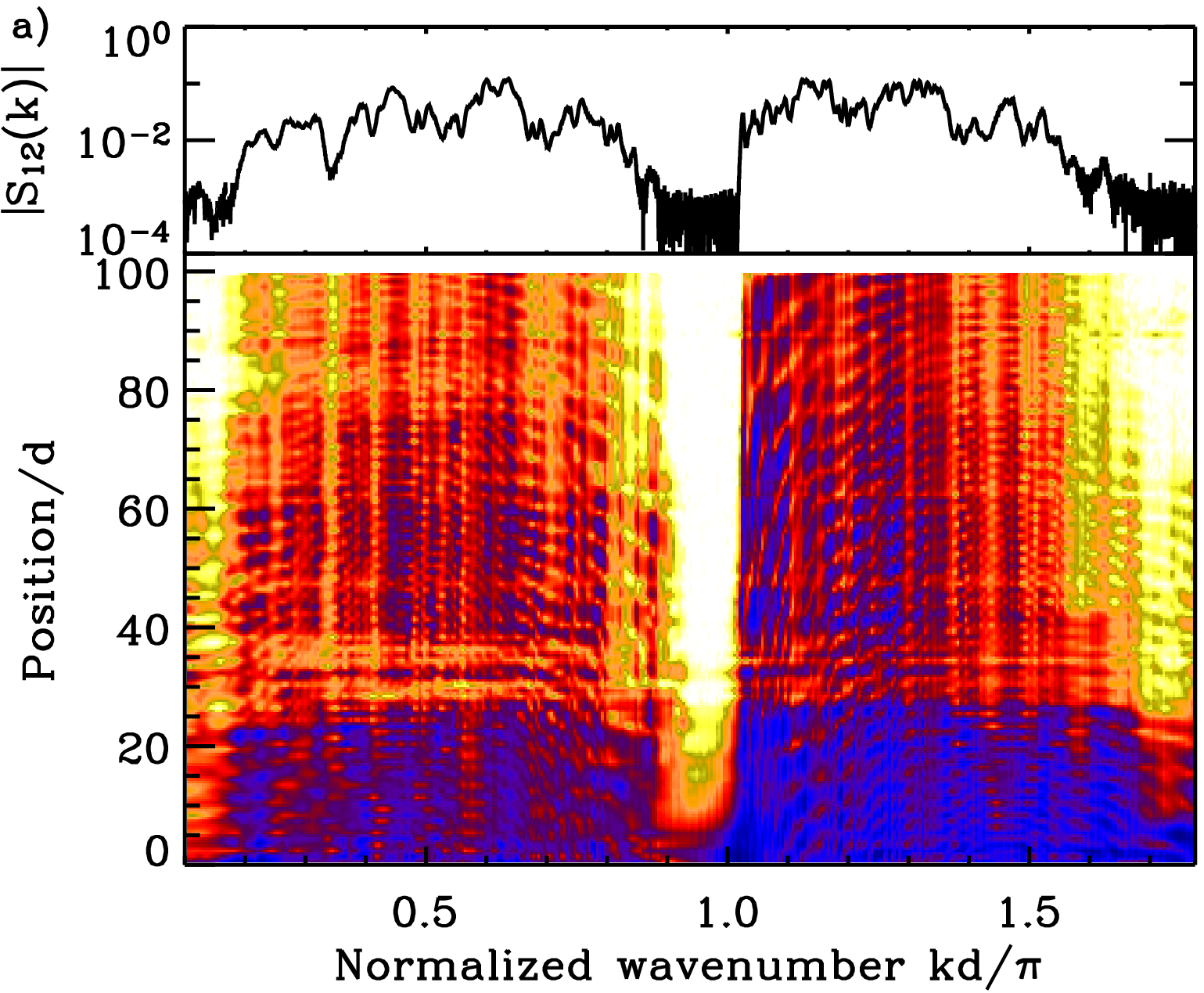}\\
\includegraphics[width=\picwidth]{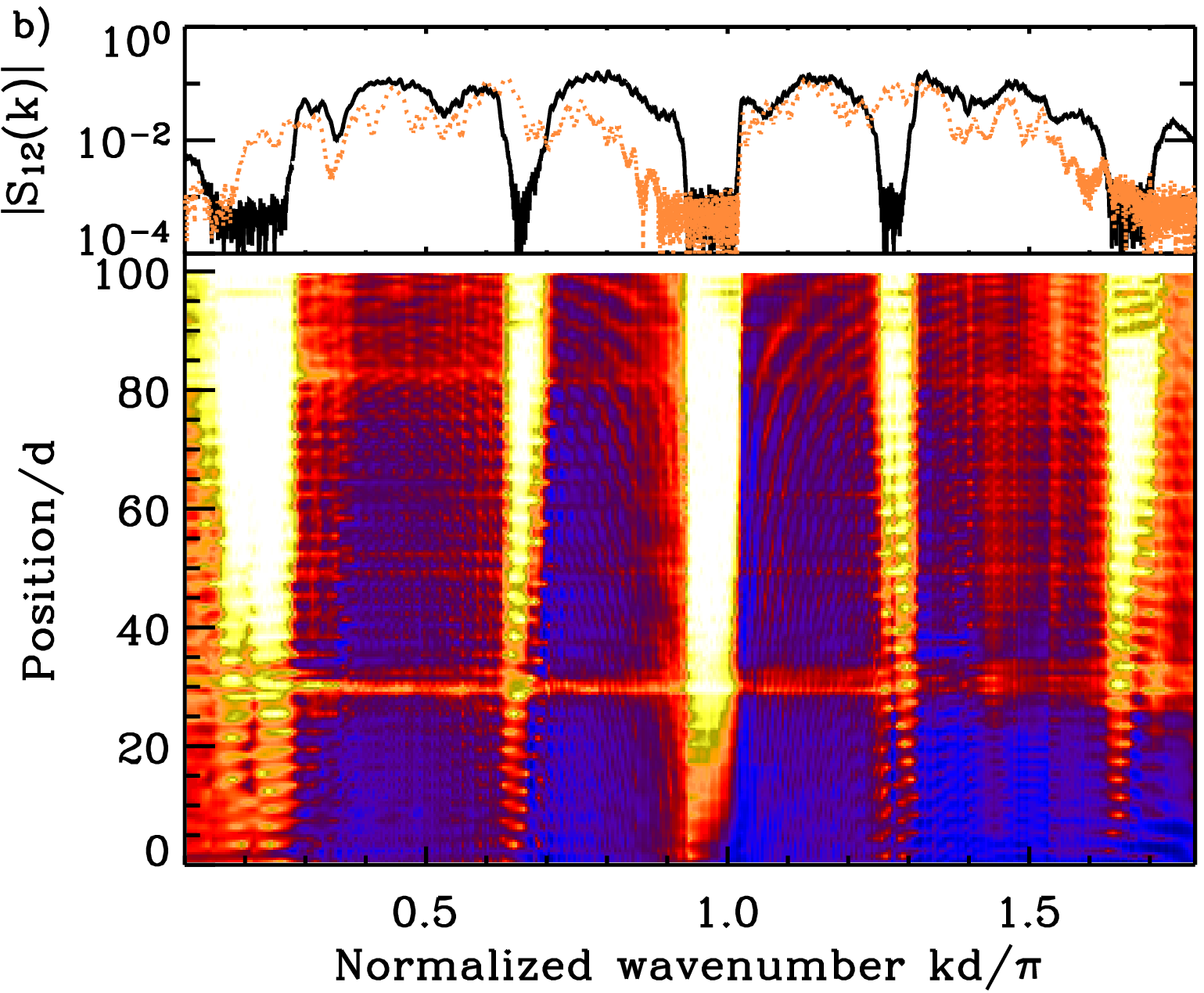}\\
\caption{\label{fig:transmission}
(Color online)
Transmission spectrum $|S_{12}(k)|$ versus position $n$ (vertical scale) of
the moving antenna is shown for uncorrelated (a) and
correlated (b) disorder. The transmission spectrum $|S_{12}|$
through the whole waveguide is shown by black solid line at the top of
each panel. For comparison, the transmission through uncorrelated disorder
is added on the top of panel (b) by a yellow dotted line.}
\end{figure}

The single-mode transmission patterns $|S_{12}|$ are shown in
Fig.~\ref{fig:transmission} for the purely random (upper) and
correlated (lower) sequence (see also Fig.~\ref{fig:disorder}) as
a function of antenna position (vertical axis) and wave number $k$
(horizonal axis). In addition, on top of each figure, we present
the dependence of transmission value $|S_{12}|$ through the whole
waveguide. As one can see, for the uncorrelated disorder there is
a gap close to the edge of the Brillouin zone $kd/\pi = 1$. It
originates from the periodic spacing between the scatterers.
For small wave numbers the transmission is small because of
the weak antenna coupling to the waveguide, whereas for high wave numbers
it is small due to large absorption. In case of the
correlated disorder there are additional gaps located at $kd/\pi
\approx 0.25, 0.65, 1.25$, and $1.65$. These gaps originate from
enhancement of localization due to long-range correlations
with $\mu_1 = 0.2$ and $\mu_2 = 0.3$ (see Eq.~(\ref{eq:corr})). It is
important that inside these gaps the transmission is practically
zero, since the localization length is reduced by a factor of $W_0
\approx 15.7$ and it becomes much less than the length of the
waveguide.

Outside the gaps the transmission is 5-10\%. This is a bit less
than within a band for a periodic arrangement. At the same time it
is larger for the correlated disorder case than in the
uncorrelated one (see fig.~\ref{fig:transmission}(b) top). Since
the relative size of the windows in which the transmission is
practically zero is much less than the band width, the integrated
transmission for the correlated disorder is also larger than the
one for the white-noise potential (see top of
fig.~\ref{fig:transmission}b)). The quantity that is conserved is
the integrated logarithmic transmission. Due to the frequency
dependence of absorption, and a noise level of $|S_{12}|^2 \approx
10^{-6}$, it is not possible to obtain the integrated logarithmic
transmission from the experiment.

\begin{figure}
\includegraphics[width=\picwidth]{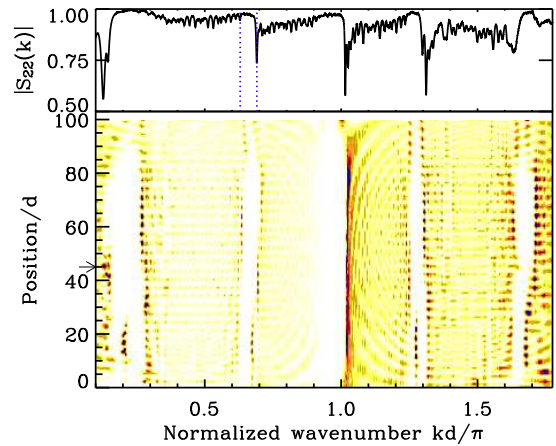}\\
\caption{\label{fig:reflection} (Color online) The same as in
Fig.~\ref{fig:transmission}(b) but for the reflection pattern
$|S_{22}(k)|$. The value $|S_{22}|$ is shown for the antenna
located between the 41st and 42nd scatterer.}
\end{figure}

\begin{figure}[h]
\includegraphics[width=\picwidth]{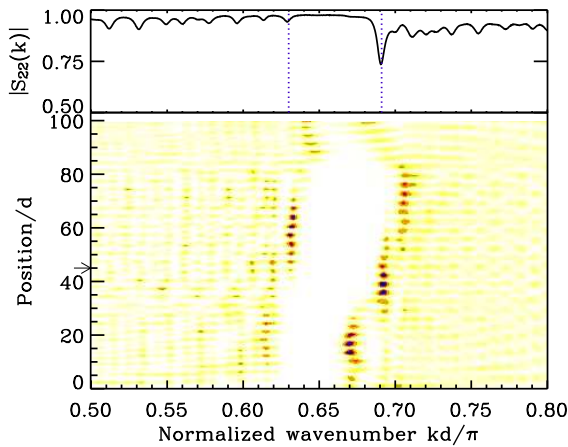}\\
\caption{\label{fig:reflectionzoom}
(Color online) Details of
Fig.~\ref{fig:reflection} for wave numbers between 0.50 and 0.80.
In this range of wave numbers one can see several enhanced localized
states. Sharp drop of $|S_{22}|$ at $kd/\pi \approx 0.691$ is due
to localized states centered at $n \approx 40$.}
\end{figure}

The emergence of the additional gaps is also seen in
Fig.~\ref{fig:reflection}, where the reflection pattern $|S_{22}|$
is shown for the correlated arrangement of scatterers. On top of the
pattern, the reflection between the 41st and 42nd scatterer is
shown. Although the localized states usually lead to strong decrease
of $|S_{22}|$, we observe close to unity reflection within the gaps
introduced by the correlations. Note that the fluctuations are
strongly suppressed in comparison to the fluctuations in the band, e.\,g.\ at $kd/\pi=0.55$.
This is even better seen in Fig.~\ref{fig:reflectionzoom} showing a narrow interval of the
wave numbers close to the correlation gap at $kd/\pi \approx 0.65$.
Several patterns with relatively high intensity of $|S_{22}|$
correspond to localized states, where the localization length is
much smaller compared to the localization length in a white-noise
arrangement. We denote these states as ``enhanced localized states".
They belong to the spectrum of the system with long-range
correlations Eq.~(\ref{eq:corr}). We observe that they are neither
evanescent modes, nor defect states \cite{lun1}. Both type of
states may in principle appear within a "band gap" if some
periodicity is introduced by the maxima of the correlation function.
For evanescent modes the reflection $S_{22}$ would be close to
1, independently of the antenna position inside the sample. We
stress that the local density of states does not change crucially by
correlations. Since each pair of scatterers brings in one state,
in the range $0.61 \le kd/\pi \le 0.71$ (shown in
Fig.~\ref{fig:reflectionzoom}) there are about 10 states, assuming
that the mean density of states is constant. Accordingly, we
observe about 8 states in Fig.~\ref{fig:reflectionzoom}.

\begin{figure}
\includegraphics[width=\picwidth]{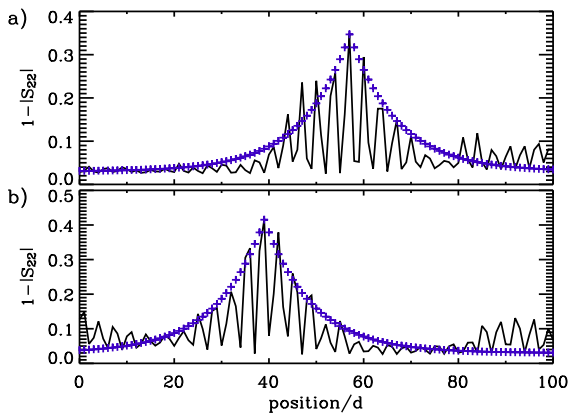}\\
\caption{\label{fig:wavefunctions} (Color online) Profiles of two
enhanced localized states at $kd/\pi \approx 0.63$ (a) and
$0.69$ (b) for the correlated sequence (marked by vertical
dotted lines in Fig.~\ref{fig:reflection} and
\ref{fig:reflectionzoom}). Graph (b) is the state responsible
for a sharp decrease of $|S_{22}|$ shown at the top of
Fig.~\ref{fig:reflectionzoom}. Crosses show an exponential
decay with a localization length of 10 scatterer spacings.}
\end{figure}

Two enhanced localized states are shown in
Fig.~\ref{fig:wavefunctions}. The quantity $1-|S_{22}|$, which is
proportional to the intensity $|\psi|^2$ of the wavefunction
\cite{kuh07b}, is plotted versus the coordinate of the moving
antenna. Exponentially localized states are clearly seen inside the
waveguide. The localization length is about 10 spacings $d$ between
the scatterers. Thus, the emergence of enhanced localized states due
to long-range correlations is found experimentally. It is highly
non-trivial that such strong enhancement of localization occurs for
relatively weak fluctuations of the potential.

We would like to stress the good agreement between theory
and experiment, in spite of the fact that {\it i)} the analytical
results are based on the analysis of the Lyapunov exponent for an
infinite sample; {\it ii)} the effect of long-range correlations is
based on the binary correlator that is correct in the first Born
approximation, {\it iii)} the number of scatterers is quite small,
{\it iv)} there is absorption of about 0.04\,dB per unit cell $d$
for the empty waveguide. Also, the potential was scaled
linearly to the micrometer screw depth, which is an approximation.
However, a strong enhancement of localization due to long-range
correlations is clearly seen, indicating that the observed effect is
robust. The method may find various applications in the design of
1D structures especially as such localized states are controlled in
the frequency space, a fact that may be important, for example, in
random lasing \cite{cao99,pat03}.

In conclusion, we performed experimental study of the effect of an
enhancement of localization, that is due to specific long-range
correlations in random potentials. Enhanced-localized states emerge
inside the single-mode microwave waveguide, within two narrow
frequency intervals. The enhancement factor (of about 16 in the
experiment) for the localization length is inversely proportional to
the width of frequency intervals. The positions of these intervals
are in a good agreement with our theoretical predictions. These
localized states appear for a quite small number of scatterers,
$N=100$, thus the theory works well far beyond the region of its
applicability.

This work was supported by the DFG via the Forschergruppe 760
'Scattering Systems with Complex Dynamics', and by the DOE grant
No.~DE-FG02-06ER46312.


\begin{thebibliography}{19}
\bibitem[{\citenamefont{Kuhl}(2007)}]{endnote}
{In fact, Eq.\protect \nobreakspace  {}(\ref {eq:norm}) is a conservation of the integrated Lyapunov exponent. This becomes clear if the function $W(2\mu )$ is replaced (using Eq. (\ref {eq:locKP})) by $l^{-1}(E)/l^{-1}_0(E)$, where $l_0(E)$ corresponds to white-noise with $W(2\mu )=1$.}
\end{thebibliography}
\end{document}